\documentclass{PoS}

\usepackage[sort&compress,numbers]{natbib}
\usepackage{graphicx}
\usepackage{amssymb}
\usepackage{amsmath}
\usepackage{bbold}
\usepackage{epstopdf}
\usepackage{float}
\usepackage{caption}
\usepackage{slashed}
\usepackage{verbatim}
\usepackage[singlelinecheck=false]{subcaption}

\DeclareGraphicsExtensions{.pdf,.png,.jpg}

\DeclareMathOperator{\tr}{tr}
\def\chpt{\raise0.4ex\hbox{$\chi$}PT}
\def\mhat{\widehat m}
\def\muhat{\widehat \mu}

\title{Phase diagram of non-degenerate Wilson and twisted mass fermions}

\ShortTitle{Phase diagram of non-degenerate Wilson and twisted mass fermions}

\author{\speaker{Derek P. Horkel}\\%
       Physics Department, University of Washington, Seattle, WA 98195-1560, USA\\
       E-mail: \email{dhorkel@uw.edu}}

\author{Stephen R. Sharpe\\%
       Physics Department, University of Washington, Seattle, WA 98195-1560, USA\\
       E-mail: \email{srsharpe@uw.edu}}

\abstract{In this talk we determine the phase diagram and pion spectrum for Wilson and twisted-mass fermions in the presence of non-degeneracy between the up and down quark and discretization errors. We find that the CP-violating phase of the continuum theory, which occurs for sufficiently large non-degeneracy, is continuously connected to the Aoki phase found in the lattice theory with degenerate quarks. Both for the Aoki and first-order scenarios, this results in a critical surface along which at least one of the pions is massless. In the pion spectrum, we focus mainly on the untwisted case, where there is competition between the effects of non-degeneracy and discretization errors. A more extensive analysis can be found in our recent paper [11].}

\FullConference{The 32nd International Symposium on Lattice Field Theory\\
		 23-28 June, 2014\\
		 Columbia University New York, NY}

\begin{document}

\section{Introduction}

Years ago, Dashen showed that for three non-degenerate quarks there is a CP-violating phase when one quark has a sufficiently negative mass ~\cite{Dashen:1970et}. This was reiterated more recently in modern language by Creutz, who showed in the framework of chiral perturbation theory (\chpt) that the transition occurs when $m_u=-m_d m_s/(m_d+m_s)$ and the chiral order parameter, $\langle\Sigma\rangle$, becomes complex~\cite{Creutz:2003xu}. The transition is second order with the neutral pion becoming massless along this phase boundary. Here we study how this phase is effected by discretization effects and how it is related to the Aoki phase of Wilson fermions~\cite{Aoki:1983qi}.

\section{Continuum Vacuum Structure from \chpt}
\label{sec:vacuum}

The lowest order (LO) chiral Lagrangian in Euclidean space-time is, for any
number of light flavors,
\begin{equation}
\mathcal{L}_\chi = \frac{f^2}{4}\tr\left[
\partial_\mu \Sigma \partial_\mu \Sigma^\dagger
-(\chi\Sigma^\dagger+\Sigma\chi^\dagger)\right]\,,
\end{equation}
where $\Sigma\in SU(N_f)$ and $\chi=2B_0 M$ (with $M$ the quark mass matrix), while $f\sim 92\;$MeV and $B_0$ are low-energy constants (LECs). The chiral order parameter $\langle \Sigma \rangle$ can be parameterized as $\exp(i \theta^a T^a)$, where in the three-flavor theory $T^a$ are the Gell-Mann matrices matrices, while in the two flavor theory $T^a = \tau^a$, the Pauli matrices. 

\begin{figure}[bt!]
\centering
\begin{subfigure}{0.49\textwidth}
\includegraphics[scale=.27]{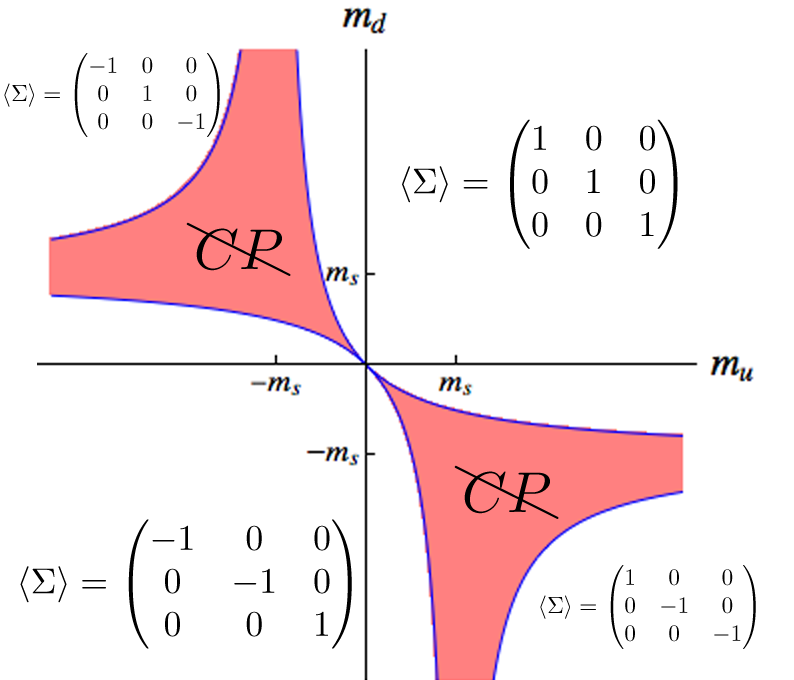}
\caption{\centering \label{fig:SU3LO} Phase diagram at lowest order in SU(3) \chpt\ 
with fixed strange quark mass.} 
\end{subfigure}
\begin{subfigure}{0.49\textwidth}
\includegraphics[scale=.25]{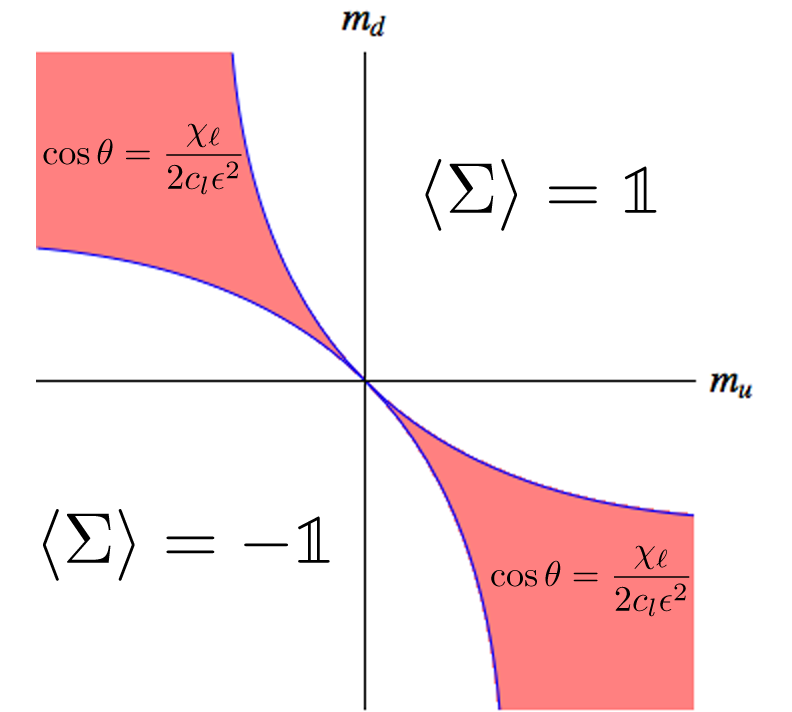}
\caption{\centering \label{fig:NLOell7} Phase diagram from SU(2) \chpt\ 
including $\ell_7$ term with $\ell_7>0$.} 
\end{subfigure}
\caption{Continuum \chpt\ phase diagram}
\end{figure}

The phase diagram of the LO $N_f=3$ theory, most extensively described by Creutz~\cite{Creutz:2003xu}, contains a CP-violating phase when either the up or down quark mass is sufficiently negative while the strange mass is fixed. The resulting phase diagram is shown in Fig.~\ref{fig:SU3LO}. The ``normal'' region, in which $\langle\Sigma\rangle=\mathbb 1$, ends at a second-order transition line along which $m_{\pi^0}$ vanishes traced out by $m_u=-m_d m_s/(m_d + m_s)$ for $m_s>0$. In this region CP is violated as $\langle \Sigma \rangle$ is now complex.

As we will be using twisted mass fermions, which work for an even number of flavors, we would like to obtain this phase diagram from two flavor \chpt. Parameterizing the condensate as $\langle\Sigma\rangle=\exp(i\theta \hat n\cdot \vec\tau)$ results in the LO potential,

\begin{equation}
\mathcal{V}_{SU(2),\,LO} 
= -\frac{f^2}{4}\tr\left[\chi\Sigma^\dagger+\Sigma\chi^\dagger\right]
= -\frac{f^2}{2}\cos{\theta}\tr[\chi] 
= -f^2 \cos{\theta}\,\chi_\ell\,
\end{equation}
where we have defined $\chi_\ell = B_0(m_u+m_d)$. At leading order the two-flavor theory only has two phases, $\theta=0$ for $\chi_\ell>0$ and $\theta=\pi$ for $\chi_\ell<0$. At LO there is no CP-violating phase for $N_f=2$.

In order to  see the CP-violating phase in the two flavor theory, we need to include a term from the NLO potential and match the two and three-flavor theories by expanding in $1/m_s$ and comparing observables in both theories. The form of the relevant NLO term is,

\begin{equation}
\mathcal{V}_{SU(2)\, NLO} \supset
\frac{\ell_7}{16}[\tr(\chi^\dagger \Sigma - \Sigma^\dagger \chi)]^2\,.
\end{equation}
Equating the $N_f=2$ NLO and the $N_f=3$ LO results for the neutral pion mass one finds~\cite{Gasser:1984gg}
\begin{equation}
\ell_7=\frac{f^2}{8B_0m_s}\,.
\label{eq:ell7match}
\end{equation}
We are neglecting the other $SU(2)$ NLO terms as they do not qualitatively change the phase diagram and are not required in matching to the LO $SU(3)$ theory as the other $\ell_i$ terms are suppressed schematically by factors of $1/\Lambda_\chi$ in opposed to $1/m_s$.

Inserting $\langle \Sigma \rangle$ into the potential yields,

\begin{equation}
\mathcal{V}_{SU(2)} = - f^2 \left( \chi_\ell \cos{\theta} 
+c_\ell \epsilon^2 n_3^2 \sin^2{\theta}\right)\,,
\label{eq:V2NLO}
\end{equation}
where we have written $\chi = \chi_\ell \mathbb 1 + \epsilon \tau_3$ with  $\epsilon= B_0(m_u-m_d)$, and  $c_\ell=\ell_7/f^2$. As the matching shows that $c_\ell>0$, minimizing the potential always aligns the condensate in the $n_3$ direction, so we can set $n_3^2=1$. Minimizing the potential now results in a CP-violating phase with $\cos\theta = \frac{\chi_\ell }{2c_\ell\epsilon^2}$, which is the minima of the potential when $\lvert \chi_\ell \rvert \leq 2c_\ell\epsilon^2$. The phase diagram is shown in Fig.~\ref{fig:NLOell7}. The pion masses in the CP-violating phase can be found by expanding $\Sigma$ about the minima, $\Sigma=\langle \Sigma\rangle\exp(i\vec \pi\cdot \vec \tau/f)$. A plot of the continuum pion masses for fixed, non-zero $\epsilon$ and varying $\chi_\ell$ is given in Fig.~\ref{fig:NLOell7}. The $SU(2)$ and the $SU(3)$ descriptions agree in the region where $m_s \gg m_{u,d}$.

\section{Including discretization effects for Wilson-like fermions}
\label{sec:disc}

The inclusion of discretization effects in \chpt\ has been thoroughly described in the literature~\cite{Sharpe:1998xm} so in the interest of brevity we will simply state the new LO operators introduced into the \chpt\ potential. With the desire for this analysis to be relevant to state-of-the-art simulations where $m_{u,d}$ is close to its physical value and lattice spacings satisfy $a^{-1} \approx 3 {\rm GeV}$, we must work in the Aoki regime, where second order discretization terms are comparable to terms linear in quark mass,

\begin{equation}
a \Lambda_{\rm QCD}^2\approx 30\,{\rm MeV} \gg m_{u,d} 
\approx a^2 \Lambda_{\rm QCD}^3 \approx 3\,{\rm MeV}\,.
\end{equation}
Absorbing the contribution linear in $a$ into the common quark mass, $\chi_\ell$, the only additional operator required at this order is the $W'$ term, such that the LO potential becomes

\begin{align}
\mathcal{V}_{a^2}&=-\frac{f^2}{4} \tr({\chi}^\dagger \Sigma + \Sigma^\dagger {\chi}) - W' [\tr(\hat A^\dagger \Sigma +   \Sigma^\dagger \hat A)]^2\\
&=- f^2 \left(\chi_\ell \cos{\theta} +w' \cos^2{\theta}\right) \, ,
\end{align}
in which $\hat A=2 W_0 a \mathbb 1$ is a spurion field and $W_0$ and $W'$ are new LECs. In the second line we inserted the parameterized condensate and defined $w'={64W'W_0^2 a^2}/{f^2}$. Extremizing the potential leads to the ``normal" $\theta=0,\pi$ phases and a CP-violating phase with $\cos{\theta}=-\frac{\chi_\ell}{2w'}$~\cite{Aoki:1983qi}. The latter occurs when, $|\chi_\ell| \leq -2w'$, and thus only for $w'<0$. 

\begin{figure}[tb!]
\centering
 \begin{subfigure}{0.49\textwidth}
\includegraphics[scale=.22]{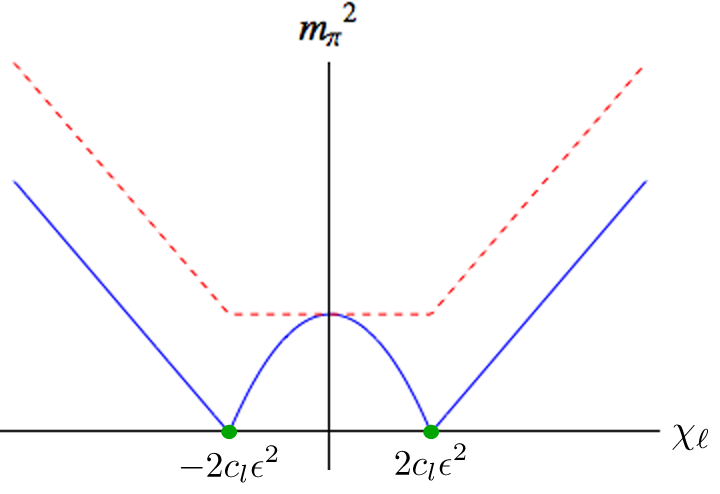}
\caption{\centering \label{fig:ContSU2}$w'=0$, $c_\ell\epsilon^2>0$}
\end{subfigure}
 \begin{subfigure}{0.49\textwidth}
\includegraphics[scale=.22]{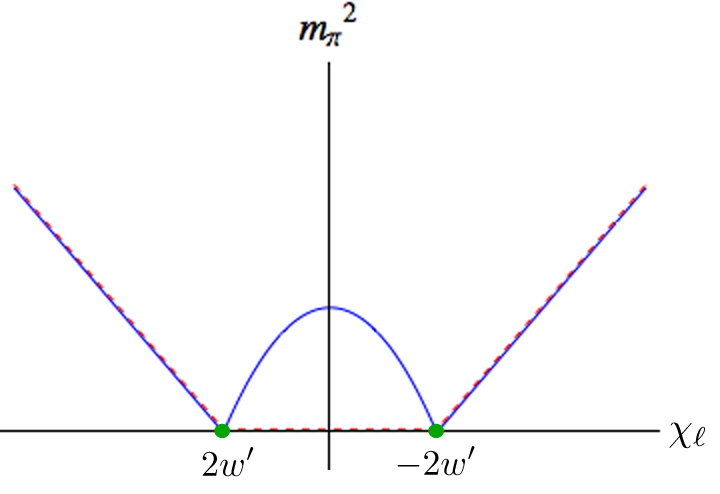}
\caption{\centering \label{fig:Aoki}$w'<0$, $c_\ell\epsilon^2= 0$}
\end{subfigure}

 \begin{subfigure}{0.49\textwidth}
\includegraphics[scale=.22]{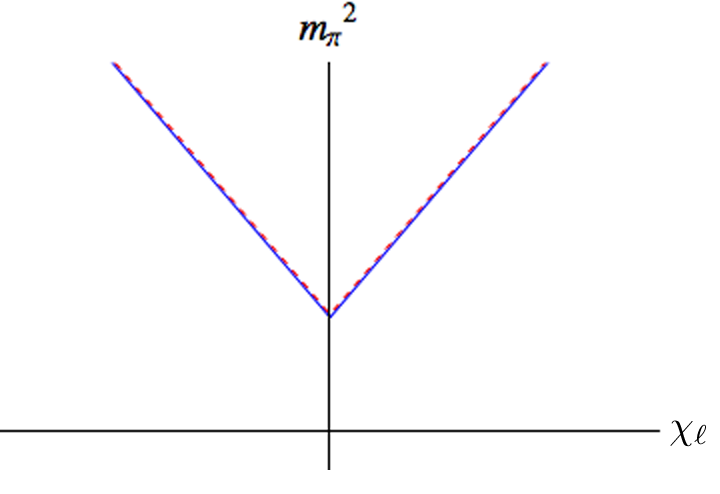}
\caption{\centering \label{fig:First}$w'>0$, $c_\ell\epsilon^2=0$}
\end{subfigure}
 \begin{subfigure}{0.49\textwidth}
\includegraphics[scale=.22]{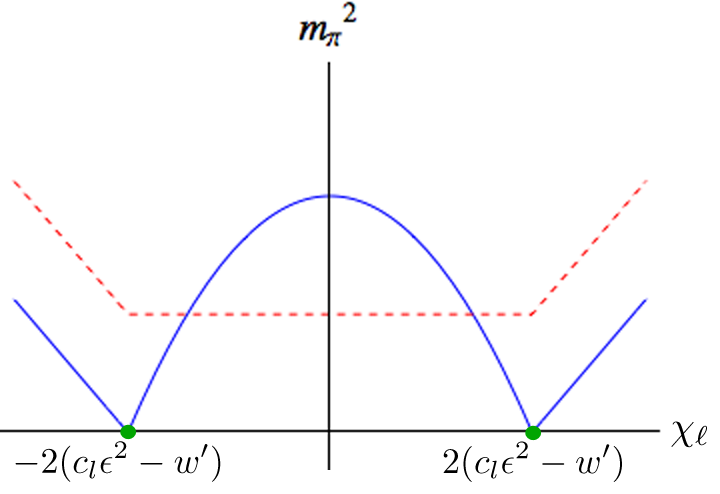}
\caption{\centering $w'<0$, $c_\ell\epsilon^2>0$}
\end{subfigure}

 \begin{subfigure}{0.49\textwidth}
\includegraphics[scale=.22]{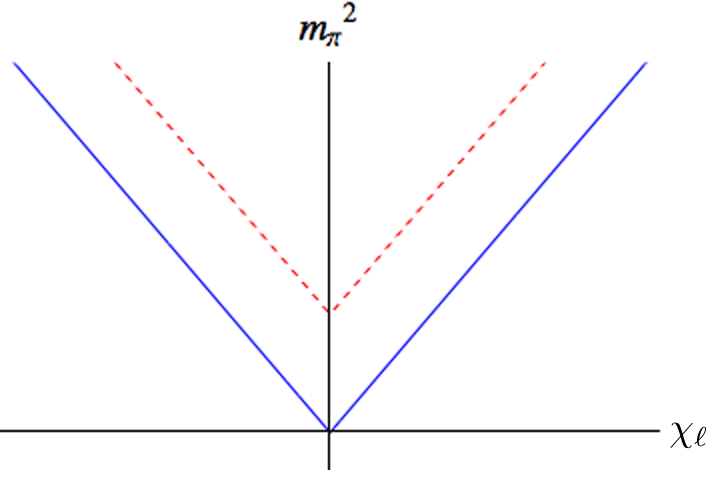}
\caption{\centering $c_\ell\epsilon^2=w'>0$}
\end{subfigure}
 \begin{subfigure}{0.49\textwidth}
\includegraphics[scale=.23]{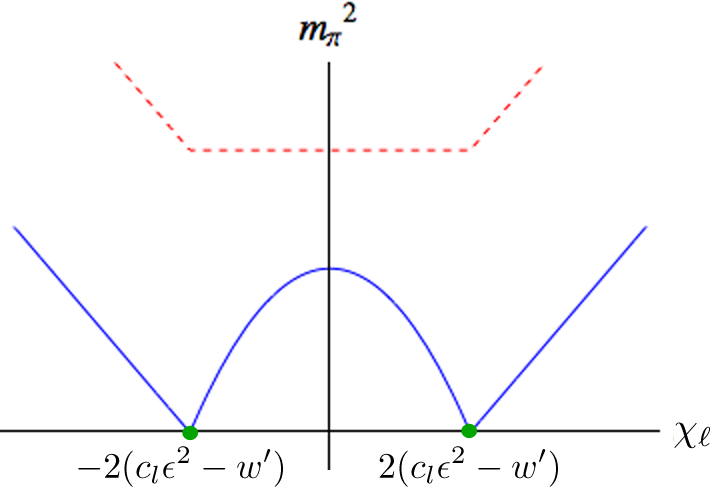}
\caption{\centering $c_\ell\epsilon^2>w'>0$}
\end{subfigure}
\caption{\label{fig:UnTwistPiMasses} 
Pion masses for untwisted Wilson
fermions including the effects of discretization 
and non-degeneracy.
$m_{\pi^0}^2$ shown by solid (blue) lines,
while $m_{\pi^\pm}^2$ by dashed (red) lines.}
\end{figure}

This results in two different scenarios: the Aoki scenario, where $w'<0$ and there is a CP-violating phase at small quark masses; and the first-order scenario, $w'>0$, where $\theta$ jumps from $0$ to $\pi$ at $\chi_\ell=0$. For the Aoki scenario, the pions are massless on either side of the phase boundary, with the charged pions being the Nambu-Goldstone bosons within the Aoki phase. For the first-order scenario there are no degenerate vacua and the pions are always massive. The pion masses for the Aoki and first-order scenario can be seen in Fig.~\ref{fig:Aoki} and~\ref{fig:First} respectively.

\begin{figure}[tb!]
\centering
\begin{subfigure}{0.49\textwidth}
\includegraphics[scale=.25]{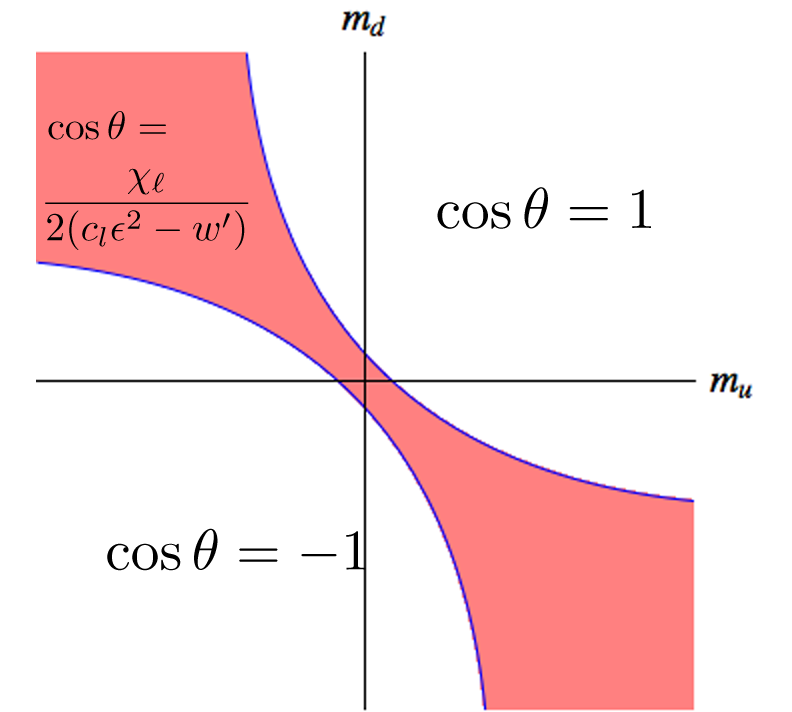}
\caption{\centering \label{fig:NLOAoki} Aoki scenario ($w'<0$).}
\end{subfigure}
\begin{subfigure}{0.49\textwidth}
\includegraphics[scale=.25]{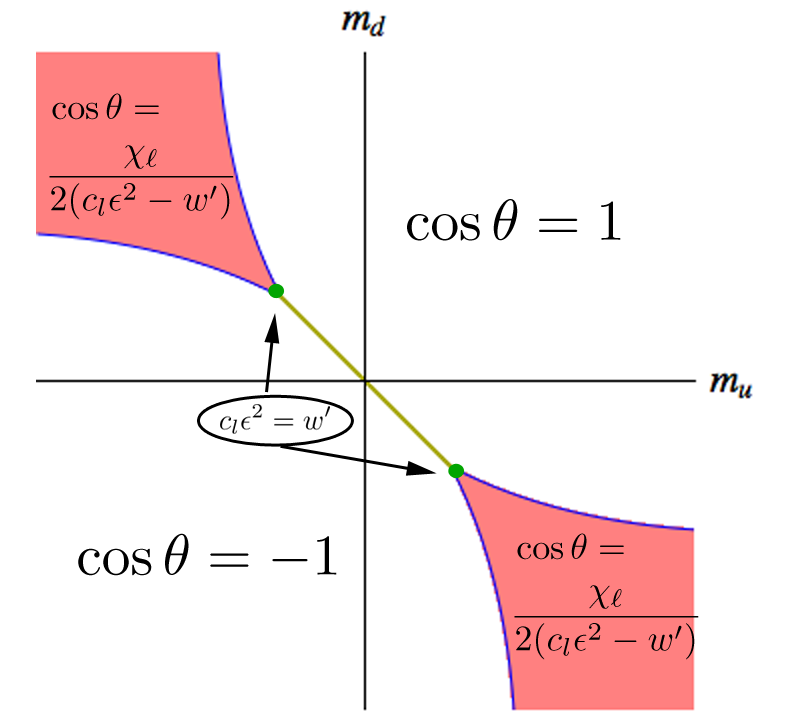}
\caption{\centering \label{fig:NLOFirst} First-order scenario ($w'>0$).}
\end{subfigure}
\caption{ Phase diagrams including effects of both
discretization ($w'\ne 0$) and non-degeneracy ($\epsilon\ne 0$).
Blue (yellow) lines indicate second (first) order transitions.}
\label{fig:untwistPhases}
\end{figure}

Including non-degeneracy along with discretization, we find the potential

\begin{equation}
- \frac{\mathcal{V}_{a^2,\ell_7}}{f^2} 
= \chi_\ell\cos{\theta} +c_\ell \epsilon^2 n_3^2
\sin^2{\theta} +w' \cos^2{\theta}\,,
\end{equation}
which is minimized for $\lvert \chi_\ell \rvert \leq 2(c_\ell \epsilon^2 -w')$ by $\cos{\theta} = \frac{\chi_\ell}{2(c_\ell \epsilon^2-w')}$. The form of this result shows that the CP-violating phase of the continuum and the Aoki phase are in fact continuously connected, as first posited by Creutz~\cite{Creutz:2014em}. The resulting phase diagram can be seen in Fig.~\ref{fig:NLOAoki} and~\ref{fig:NLOFirst} for the Aoki and first-order scenarios respectively. The resulting pion masses, seen in Fig.~\ref{fig:UnTwistPiMasses}, show that the neutral pion mass traces out the boundaries of the second order phase transition, while the charged pions are massive in the presence of non-degeneracy.

\section{Twisted-mass fermions at maximal twist}
\label{sec:twist}

\begin{figure}[tb!]
\centering
 \begin{subfigure}{0.49\textwidth}
 \centering
 \includegraphics[scale=.28]{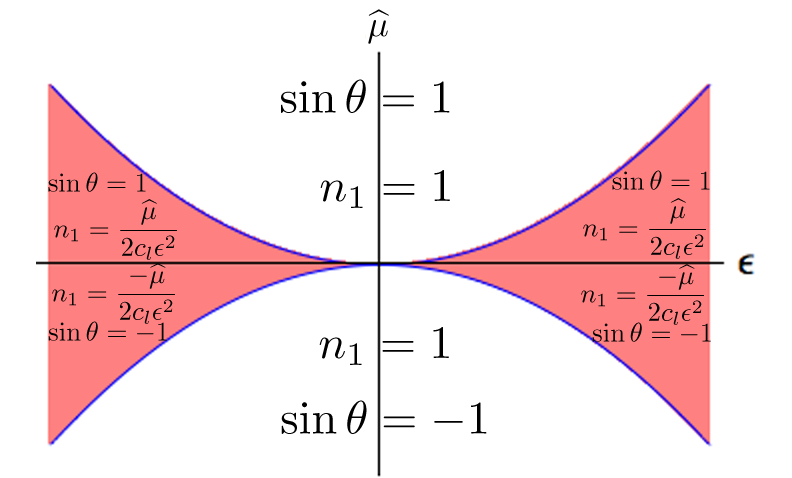}
\caption{\centering Aoki scenario or continuum ($w'\leq0$)}
\end{subfigure}
 \begin{subfigure}{0.49\textwidth}
 \centering
 \includegraphics[scale=.28]{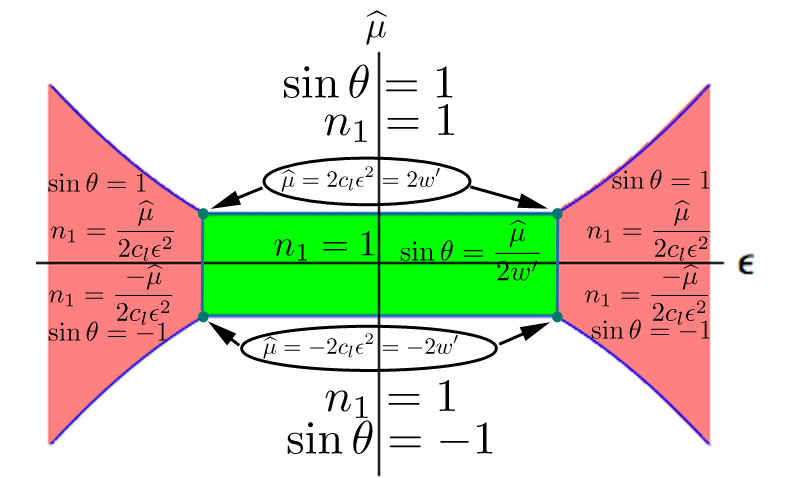}
\caption{\centering First-order scenario ($w'>0$)}
\end{subfigure}
\caption{\label{fig:maxTwistPhases} 
Phase diagrams at maximum twist ($\mhat=0$).}
\end{figure}

We now extend the previous results to twisted-mass fermions~\cite{Frezzotti:2000nk} at maximal twist. We focus on maximal twist as it has the property of automatic ${\cal O}(a)$ improvement~\cite{Frezzotti:2003ni}. We choose the twist to be in a direction orthogonal to $\tau_3$ such that the $\epsilon_\ell$ term unchanged. In the continuum this is a convenience, but when working with on a lattice with a Wilson term, it is mandatory to twist in a direction orthogonal to $\tau_3$ in order to keep the fermion determinant real~\cite{Frezzotti:2003xj}. By convention, this direction is chosen to be $\tau_1$. The rescaled mass matrix that enters \chpt\ is now ~\cite{Sharpe:2004ps},
\begin{equation}
\chi=\chi_\ell e^{i \tau_1 \pi/2} +\epsilon \tau_3 = 
i \chi_\ell \tau_1  +\epsilon \tau_3 = 
i \muhat\tau_1 +\epsilon \tau_3\,.
\label{eq:twistedchi}
\end{equation}
No new operators are introduced at this order by adding a twisted mass and $\mathcal{V}$ becomes,
\begin{equation}
- \frac{\mathcal{V}_{a^2,\ell_7}}{f^2} = 
\muhat n_1 \sin{\theta}
+c_\ell \epsilon^2 n_3^2 \sin^2{\theta} +w'\cos^2{\theta}\,.
\label{eq:Vtwist}
\end{equation}
The condensate can now point in either the $n_1$ or $n_3$ direction. The competition between twist and non-degeneracy results in a phase diagram which is identical to the continuum result in the Aoki scenario and has an additional phase for the first-order scenario, as seen in Fig.~\ref{fig:maxTwistPhases}. As the continuous $SU(2)$ flavor symmetry of the  lattice theory is completely broken at maximal twist, in general all three pion masses differ. More details of the the calculation of the pion masses and the symmetries involved can be found in Ref.~\cite{Horkel:2014nha}.

\section{Conclusions and extensions}
\label{sec:concl}

In this work we have studied how using non-degenerate up and down quarks changes the phase structure due to competition between quark mass and discretization effects. We find that the continuum CP-violating phase is continuously connected to the Aoki phase previously studied for degenerate quarks. We also find that discretization effects can move the theory with physical quark masses closer to, or even into, unphysical phases. This occurs in the Aoki scenario for untwisted Wilson quarks, and in the first-order scenario for maximally twisted quarks. Our overall message is that a complicated phase structure lies in the vicinity of the physical point and simulations should be careful to avoid unphysical phases.
More details, such as the pion masses at maximal twist, the critical manifold at arbitrary twist and $\mathcal{O}(m a \sim a^3)$ corrections are presented in our full paper~\cite{Horkel:2014nha}.
In this work we have not included electromagnetic effects. In the pion sector, these lead to isospin breaking that is generically larger than that from quark non-degeneracy.
We will discuss the impact of electromagnetism in an upcoming work,
building upon the recent analysis of Ref.~\cite{Golterman:2014yha} and ~\cite{deDivitiis:2013xla}. One key result is that the untwisted phase structure is unchanged but the question of how to include electromagnetism when working with non-zero twist is more involved.

\section*{Acknowledgments}
This work was supported in part by the United States Department of Energy 
grants DE-FG02-96ER40956 and DE-SC0011637.

\bibliography{ref}{}
\bibliographystyle{apsrev4-1.bst}

\end{document}